# Effect of inter-layer spin diffusion on skyrmion motion in magnetic multilayers

Serban Lepadatu[*]

*Jeremiah Horrocks Institute for Mathematics, Physics and Astronomy, University of Central Lancashire, Preston PR1 2HE, U.K.*

Abstract

It is well known that skyrmions can be driven using spin-orbit torques due to the spin-Hall effect. Here we show an additional contribution in multilayered stacks arises from vertical spin currents due to inter-layer diffusion of a spin accumulation generated at a skyrmion. This additional interfacial spin torque is similar in form to the in-plane spin transfer torque, but is significantly enhanced in ultra-thin films and acts in the opposite direction to the electron flow. The combination of this diffusive spin torque and the spin-orbit torque results in skyrmion motion which helps to explain the observation of small skyrmion Hall angles even with moderate magnetisation damping values. Further, the effect of material imperfections on threshold currents and skyrmion Hall angle is also investigated. Topographical surface roughness, as small as a single monolayer variation, is shown to be an important contributing factor in ultra-thin films, resulting in good agreement with experimental observations.

[*] SLepadatu@uclan.ac.uk

I. Introduction

Skyrmions are topologically protected particle-like magnetic textures [1], which are of great interest for potential technological applications. Skyrmions have been observed in materials with broken inversion symmetry and stabilised at room temperature through the Dzyaloshinskii-Moriya interaction (DMI) [2-5]. As carriers of information it is important to effectively move, as well as detect, skyrmions using electrical signals and readout. To this end, recent experiments have revealed the fascinating physics behind the interaction of skyrmions with spin currents. Current-induced skyrmion movement was demonstrated at room temperature in a number of recent experiments on ultra-thin multilayered stacks [6-12], whilst electrical readout is made possible through the discrete Hall resistivity [13]. The principal source of spin currents in these devices is the spin-Hall effect (SHE), which converts a charge current flowing in the plane into transverse pure spin currents. The resultant spin-orbit torque (SOT) gives rise to skyrmion motion, with direction set by the charge current direction, as well as skyrmion chirality [14]. The skyrmion Hall effect, whereby the direction of skyrmion movement deviates from that of the charge current direction due to the Magnus force, was also demonstrated [8-10]. Non-zero skyrmion Hall angles present an obstacle to realising skyrmion-based spintronics devices. Strategies to reduce the skyrmion Hall angle to zero have been proposed, using antiferromagnetically exchange-coupled bilayer systems [15], as well as using skyrmionium magnetic textures [16].

Due to spin precession of spin-polarised electrons flowing through a magnetic texture, a spin accumulation is generated at magnetisation texture gradients resulting in adiabatic and non-adiabatic spin transfer torques (STT) [17,18]. Furthermore spin diffusion was also shown to play a role, resulting in modified diffusive spin torques when considering two-dimensional magnetic textures [19]. On the other hand vertical spin currents have been shown to play a more important role in driving skyrmions in nanostructures [20], whilst the importance of interfaces and interface-generated spin currents has also been recognised [21]. Here we show, using micromagnetics modelling coupled with a self-consistent spin transport solver in multilayers, that the spin accumulation generated at the magnetisation gradients of a skyrmion results in additional vertical spin currents due to spin diffusion in adjacent non-magnetic layers. These diffusive spin currents result in additional interfacial spin torques which can be comparable to the SOT, significantly reducing the calculated skyrmion Hall angle even for small magnetisation damping values. In experiments it was found the skyrmion Hall angle strongly depends on the skyrmion velocity, evidencing the important role material imperfections play [8-10]. Using the self-consistent spin transport solver we also study the effect of SOTs and inter-layer diffusion in the presence of magnetic defects, as well as topographical surface roughness. In particular surface roughness is shown to result in strong confining potentials, resulting in a dependence of the skyrmion Hall angle with driving current, as well as threshold current densities comparable to those found in experiments. These results may indicate an alternative method of designing devices with zero skyrmion Hall angle, by purposely creating surface confining potentials.

## II. Spin Transport Model

Spin torques included in the magnetisation dynamics equation can be computed self-consistently using a drift-diffusion model [22,23]. Within this model the charge and spin current densities are given as:

$$\mathbf{J}_C = \sigma\mathbf{E} + \theta_{SHA} D_e \frac{e}{\mu_B} \nabla \times \mathbf{S} \tag{1}$$

$$\mathbf{J}_S = -\frac{\mu_B}{e} P\sigma\mathbf{E} \otimes \mathbf{m} - D_e \nabla\mathbf{S} + \theta_{SHA} \frac{\mu_B}{e} \boldsymbol{\varepsilon}\sigma\mathbf{E} \tag{2}$$

Here $\mathbf{J}_S$ is a rank-2 tensor such that $\mathbf{J}_{Sij}$ signifies the flow of the $j$ component of spin polarisation in the direction $i$. Equation (2) contains contributions due to i) drift included in ferromagnetic (F) layers, where $P$ is the current spin-polarisation and $\sigma$ the electrical conductivity, ii) diffusion, where $D_e$ is the electron diffusion constant, and iii) spin-Hall effect, included in non-magnetic (N) layers, where $\theta_{SHA}$ is the spin-Hall angle and $\boldsymbol{\varepsilon}$ is the rank-3 unit antisymmetric tensor. The inverse spin-Hall effect is included for completeness as a contribution in Equation (1). The spin accumulation, $\mathbf{S}$, satisfies the equation of motion:

$$\frac{\partial \mathbf{S}}{\partial t} = -\nabla \cdot \mathbf{J}_S - D_e \left( \frac{\mathbf{S}}{\lambda_{sf}^2} + \frac{\mathbf{S}\times\mathbf{m}}{\lambda_J^2} + \frac{\mathbf{m}\times(\mathbf{S}\times\mathbf{m})}{\lambda_\varphi^2} \right) \tag{3}$$

Here $\lambda_{sf}$ is the spin-flip length which governs the decay of spin accumulation. In F layers the decay of transverse components of **S** are governed by the exchange rotation length $\lambda_J$, and the spin dephasing length $\lambda_\varphi$. Solving Equations (1)-(3), we obtain a Poisson-type equation for the steady-state spin accumulation:

$$\nabla^2 \mathbf{S} = -\frac{P}{D_e}\frac{\mu_B}{e}(\mathbf{J}_C.\nabla)\mathbf{m} + \frac{\theta_{SHA}}{D_e}\frac{\mu_B}{e}\nabla.(\varepsilon\mathbf{J}_C) + \frac{\mathbf{S}}{\lambda_{sf}^2} + \frac{\mathbf{S}\times\mathbf{m}}{\lambda_J^2} + \frac{\mathbf{m}\times(\mathbf{S}\times\mathbf{m})}{\lambda_\varphi^2} \qquad (4)$$

Thus for each magnetisation configuration **m** the resulting spin accumulation is obtained by solving Equation (4). This is justified since **m** and **S** vary on very different timescales (ps vs fs respectively). In Equation (1) we have the usual relation **E** = -∇*V*. For boundaries containing an electrode with a fixed potential, differential operators applied to *V* use a Dirichlet boundary condition. For other external boundaries we require both the charge and spin currents to be zero in the direction normal to the boundary, i.e. **J**$_C$.**n** = 0 and **J**$_S$.**n** = 0 [24]. This results in the following non-homogeneous Neumann boundary conditions:

$$\nabla V.\mathbf{n} = \theta_{SHA}\frac{D_e}{\sigma}\frac{e}{\mu_B}(\nabla\times\mathbf{S})\mathbf{n}$$

$$(\nabla\mathbf{S})\mathbf{n} = \theta_{SHA}\frac{\sigma}{D_e}\frac{\mu_B}{e}(\varepsilon\mathbf{E})\mathbf{n} \qquad (5)$$

At the interface between two N layers we obtain composite media boundary conditions for *V* and **S** by requiring both a potential and associated flux to be continuous in the direction normal to the interface, i.e. *V* and **J**$_C$, and **S** and **J**$_S$ respectively. At an N/F interface we do not assume such continuity, but instead model the absorption of transverse spin components using the spin-mixing conductance [25]:

$$\mathbf{J}_C.\mathbf{n}\big|_N = \mathbf{J}_C.\mathbf{n}\big|_F = -(G^\uparrow + G^\downarrow)\Delta V + (G^\uparrow - G^\downarrow)\Delta\mathbf{V}_S.\mathbf{m}$$

$$\mathbf{J}_S.\mathbf{n}\big|_N - \mathbf{J}_S.\mathbf{n}\big|_F = \frac{2\mu_B}{e}\left[\text{Re}\{G^{\uparrow\downarrow}\}\mathbf{m}\times(\mathbf{m}\times\Delta\mathbf{V}_S) + \text{Im}\{G^{\uparrow\downarrow}\}\mathbf{m}\times\Delta\mathbf{V}_S\right] \qquad (6)$$

$$\mathbf{J}_S.\mathbf{n}\big|_F = \frac{\mu_B}{e}\left[-(G^\uparrow + G^\downarrow)(\Delta\mathbf{V}_S.\mathbf{m})\mathbf{m} + (G^\uparrow - G^\downarrow)\Delta V\mathbf{m}\right]$$

Here $\Delta V$ is the potential drop across the N/F interface ($\Delta V = V_F - V_N$) and $\Delta \mathbf{V}_S$ is the spin chemical potential drop, where $\mathbf{V}_S = (D_e/\sigma)(e/\mu_B)\mathbf{S}$, and $G^{\uparrow}$, $G^{\downarrow}$ are interface conductances for the majority and minority spin carriers respectively. The transverse spin current absorbed at the N/F interface results in a torque on the magnetisation as a consequence of conservation of total spin angular momentum. If the F layer has thickness $d_F$, this interfacial torque is obtained as:

$$\mathbf{T}_S = \frac{g\mu_B}{ed_F}\left[\mathrm{Re}\{G^{\uparrow\downarrow}\}\mathbf{m}\times(\mathbf{m}\times\Delta\mathbf{V}_S) + \mathrm{Im}\{G^{\uparrow\downarrow}\}\mathbf{m}\times\Delta\mathbf{V}_S\right] \tag{7}$$

In the equation of motion for **m**, this interfacial torque is included as:

$$\frac{\partial \mathbf{m}}{\partial t} = -\gamma \mathbf{m}\times\mathbf{H}_{eff} + \alpha\mathbf{m}\times\frac{\partial \mathbf{m}}{\partial t} + \frac{1}{M_S}\mathbf{T}_S \tag{8}$$

Here $\gamma = \mu_0 g_{rel}|\gamma_e|$, where $\gamma_e = -g\mu_B/\hbar$ is the electron gyromagnetic ratio, $g_{rel}$ is a relative g-factor, and $M_s$ is the saturation magnetisation. Using Equation (6) we can also include spin pumping on the N side of the equation as [26]:

$$\mathbf{J}_S^{pump} = \frac{\mu_B \hbar}{e^2}\left[\mathrm{Re}\{G^{\uparrow\downarrow}\}\mathbf{m}\times\frac{\partial \mathbf{m}}{\partial t} + \mathrm{Im}\{G^{\uparrow\downarrow}\}\frac{\partial \mathbf{m}}{\partial t}\right] \tag{9}$$

For an N/F interface with current in the plane, if diffusion effects are negligible, the drift-diffusion equations may be solved analytically to obtain the resulting interfacial spin torques due to SHE [23]. These are given as a combination of damping-like and field-like spin-orbit torques as:

$$\mathbf{T}_{SOT} = \theta_{SHA,eff}\frac{\mu_B}{e}\frac{|J_c|}{d_F}[\mathbf{m}\times(\mathbf{m}\times\mathbf{p}) + r_G \mathbf{m}\times\mathbf{p}] \tag{10}$$

Here $\mathbf{p} = \mathbf{z} \times \mathbf{e_{Jc}}$, where $\mathbf{e_{Jc}}$ is the charge current direction. The effective spin-Hall angle is given by:

$$\theta_{SHA,eff} = \theta_{SHA}\left(1 - \frac{1}{cosh(d_N/\lambda_{sf}^N)}\right)\frac{N_\lambda \text{Re}\{\tilde{G}\} + |\tilde{G}|^2}{(N_\lambda + \text{Re}\{\tilde{G}\})^2 + \text{Im}\{\tilde{G}\}^2}, \qquad (11)$$

where $N_\lambda = tanh(d_N/\lambda_{sf}^N)/\lambda_{sf}^N$, and $\tilde{G} = 2G^{\uparrow\downarrow}/\sigma_N$. The field-like torque coefficient is given by $r_G = N_\lambda \text{Im}\{\tilde{G}\}/(N_\lambda \text{Re}\{\tilde{G}\} + |\tilde{G}|^2)$. Note, in the limit of transparent interface ($\text{Re}\{G^{\uparrow\downarrow}\} \to \infty$) the field-like torque tends to zero, and if further $d_N \gg \lambda_{sf}^N$ we obtain the approximation $\theta_{SHA,eff} \cong \theta_{SHA}$.

With N/F multilayers another important source of vertical spin currents, resulting in an interfacial spin torque contribution, is due to N/F inter-layer diffusion of a spin accumulation generated in the F layer at spatial gradients in the magnetisation texture, e.g. a skyrmion. This is in some ways similar to the in-plane STT arising in the F layer alone [17,18], but in ultra-thin films the inter-layer diffusion results in much stronger spin torques partly due to the inverse dependence on $d_F$. It can be shown this additional interfacial spin torque is given by (see Supplementary Material for derivation):

$$\mathbf{T}_{Diff.} = -\left[(\mathbf{u}_\perp \cdot \nabla)\mathbf{M} - \frac{\beta_\perp}{M_S}\mathbf{M} \times (\mathbf{u}_\perp \cdot \nabla)\mathbf{M}\right] \qquad (12)$$

This interfacial spin torque has a very similar form to the well-known Zhang-Li STT, with the exception it acts in the opposite direction, i.e. results in motion along the current direction, and the spin-drift velocity and non-adiabaticity parameters are replaced by effective perpendicular spin-drift velocity and perpendicular non-adiabaticity parameters. In particular the perpendicular spin-drift velocity is given by:

$$\mathbf{u}_\perp = \mathbf{J}_C \frac{P_\perp g \mu_B}{2eM_S}, \qquad (13)$$

where $P_\perp$ is an effective perpendicular spin polarisation parameter. These parameters are not dependent on a single material alone, but are effective parameters for the entire multilayered stack.

## III. Spin Torques in Multilayers

Current-induced Néel skyrmion movement has been observed in a number of ultra-thin multilayered stacks, including Ta/CoFeB/TaO$_x$ [6,9], [Pt/Co/Ta]$_x$ [7], Ta/[Pt/Ir/Co]$_x$/Pt [8], [Pt/CoFeB/MgO]$_x$ [7,10], [Pt/GdFeCo/MgO]$_x$ [11], and symmetric bilayer stacks [12]. To study the effect of the spin torques described in the previous section on skyrmion motion, a multilayered disk geometry was chosen, with the structure [Pt (3 nm)\Co (1 nm)\Ta (4 nm)]$_x$, which has been well-characterised experimentally [7,27,28,29].

**Figure 1** – Multilayered [Pt/Co/Ta]$_6$ 320 nm diameter disk with a Néel skyrmion, showing the charge current density injected through the bottom Pt layer. The images below show the z-direction spin current density in a Pt layer, as generated due to inter-layer spin diffusion, as well as the resulting spin torque on a Co layer.

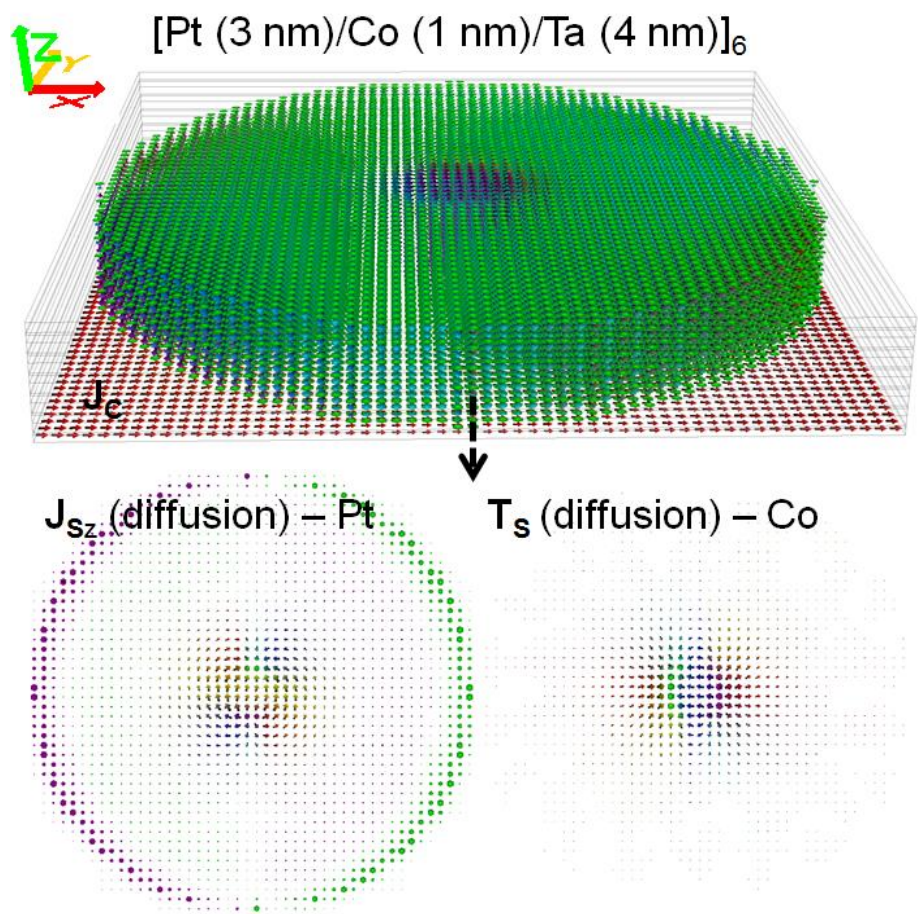

The disk geometry was chosen so the movement of skyrmions in different directions under the various spin torques is not affected by shape anisotropy. The studied geometry is shown in Figure 1 for a repetition of 6 Pt/Co/Ta stacks (x = 6). The bottom Pt layer was extended and a current applied to the structure through electrodes on its x-axis ends. This configuration ensures that, apart from the edges of the disks, the current density is approximately uniform (less than 2% variation in the region where skyrmion motion is simulated). A table with the full list of material parameters used is given in the Methods section.

First we investigate the effect of spin torques on a skyrmion in a single Pt/Co/Ta stack for fixed current density, for different chiralities and topological charges. Using a fixed out-of-plane field $|H_z| = 15$ kA/m the skyrmion diameter is fixed to 60 nm, similar to that observed experimentally [7]. The results are shown in Figure 2. For the spin torque obtained with the self-consistent spin transport solver we see two distinct contributions. In addition to the spin torque due to SHE alone, namely SOT, an equally important contribution is obtained due to inter-layer spin diffusion. To demonstrate this, skyrmions have been driven with and without SHE contribution. Without SHE ($\theta_{SHA} = 0$ in both Pt and Ta) the only torque acting on the skyrmions is due to inter-layer spin diffusion, as seen from the good agreement between spin transport solver computations with $\theta_{SHA} = 0$, and simulations using the LLG equation complemented by the diffusive spin torque in Equation (12). The vertical spin current due to diffusion is shown in the Pt layer in Figure 1, as well as the resulting interfacial spin torque acting on the skyrmion. Note, in this work we didn't consider Zhang-Li in-plane STTs since their effect is much smaller in ultra-thin films compared to interfacial spin torques [20].

For Equation (12) we find $P_\perp = 0.87$ and $\beta_\perp = -0.13$ for the interfacial spin torque. The large effective perpendicular spin polarisation and perpendicular non-adiabaticity parameters result in a total spin torque comparable to the SOT. In contrast to SOTs however, the direction of motion is opposite to the electron flow in all cases, for both D < 0 and D > 0. Further, by subtracting results obtained using the spin transport solver for $\theta_{SHA} \neq 0$ and $\theta_{SHA} = 0$ we obtain a good agreement with simulations using the LLG equation complemented by the SOT in Equation (10). As expected, with the SOT alone the direction of motion depends on the sign of the DMI. However, when inter-layer spin diffusion is taken into account, the overall effect is for skyrmion motion opposing the flow of electrons in all cases.

**Figure 2** – Néel skyrmion motion in a Pt/Co/Ta disk shown for $J_C = \pm 1.3 \times 10^{11}$ A/m$^2$, $D = \pm 1.5$ mJ/m$^2$ and different skyrmion core orientations: into the plane for (a), (c) with applied field $H_z$ = 15 kA/m, out of the plane for (b), (d) with $H_z$ = -15 kA/m. The skyrmion motion is shown for the self-consistent spin transport solver (blue lines), spin transport solver including spin pumping (green lines), spin transport solver but without SHE, thus with vertical spin currents due to inter-layer spin diffusion only (red squares). The latter is compared with simulations using the analytical form of the diffusive spin torque (grey lines) – Equation (12). The difference between spin transport solver results for cases with and without SHE is shown as magenta circles. This is compared with results using the analytical form of the SOT (black lines) – Equation (10) with the damping-like component only.

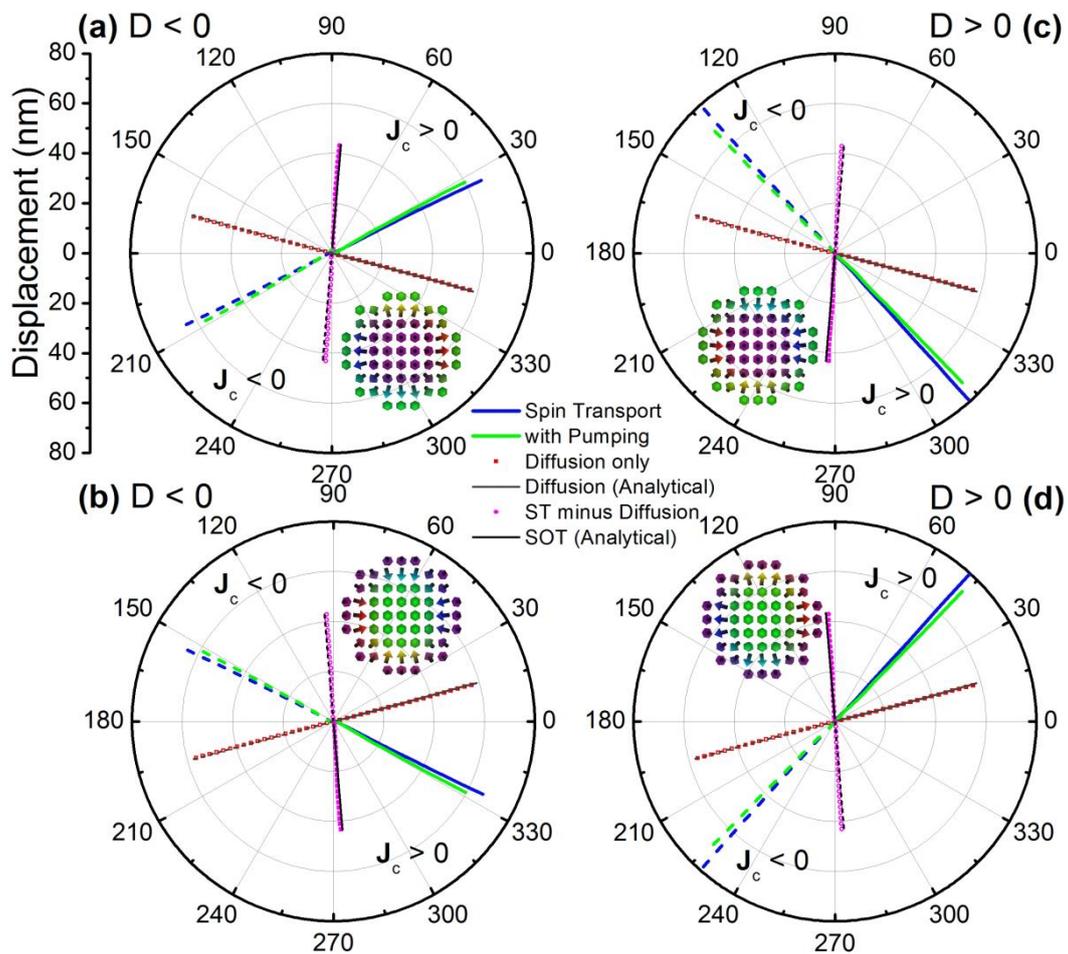

Experimental investigations of current-induced skyrmion movement have revealed skyrmion movement in the direction opposing the flow of electrons [6-12], and these results have been analysed principally based on the SOT due to SHE. We show here however, inter-layer spin diffusion could also have a significant effect and should be considered when analysing skyrmion motion. The implications are both qualitative and quantitative. Since the skyrmion motion direction due to the diffusive spin torque is always in the direction opposing the flow of electrons, if inter-layer spin diffusion is significant, the exact topology of skyrmions cannot be determined purely based on observing their motion direction (with or against the electron flow). Quantitatively, whilst the skyrmion velocities are not greatly affected by inclusion of the diffusive spin torque, due to the nearly orthogonal skyrmion movement directions under these two torques respectively, the skyrmion Hall angle obtained varies markedly and could have significant implications in explaining experimental results.

The Onsager reciprocal process to absorption of transverse spin currents is the generation of spin currents via dynamical magnetisation processes, known as spin pumping [26] – see Equation (9). The effect of spin pumping on magnetisation precession is an increase in the effective magnetisation damping. As expected from the Thiele equation [30], larger damping values should result in reduced skyrmion velocities, thus it is interesting to observe its effect on skyrmion motion. First we keep the current density fixed, and later analyse the effect of varying the current density. With spin pumping enabled in the spin transport solver, a spin drag effect is observed, resulting in a slight reduction in velocity as shown in Figure 2, as well as a slight deviation of the skyrmion path. This effect is quite small however and could be ignored in simpler simulations using only the analytical form of the SOT and diffusive spin torque.

**Figure 3** – Effect of higher damping values on skyrmion motion under various spin torques for $J_C = 1.3 \times 10^{11}$ A/m$^2$, $H_z = 15$ kA/m, and $D = \pm1.5$ mJ/m$^2$.

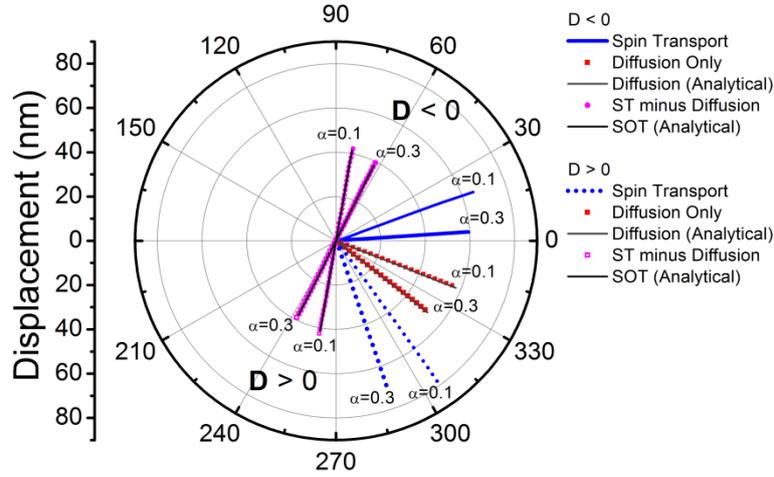

The results discussed thus far used an intrinsic damping value in Co of 0.03, as obtained using ferromagnetic resonance measurements in ultra-thin films [41]. On the other hand, much higher damping values of up to 0.3 have been obtained in Pt/Co/Pt films from magnetic domain-wall motion experiments [42]. Increasing damping results in a reduced skyrmion velocity as expected, however more significantly the direction of motion is strongly affected, resulting in a clockwise rotation of the skyrmion paths with increasing damping as seen in Figure 3. This holds for both the SOT and diffusive spin torque, and again a good agreement is obtained between the spin transport solver results and simulations using the analytical forms of the SOT and diffusive spin torque. Small skyrmion Hall angles have been observed experimentally [9,10]. Based on the SOT alone the skyrmion Hall angle is inversely dependent on the magnetisation damping [31], namely $\tan\theta_{SkH} \propto 1/\alpha$, and as noted in [10] the experimentally observed skyrmion Hall angles are smaller than those obtained using micromagnetics modelling with the SOT alone, under realistic parameters. As shown in [9] disorder plays a very significant effect on the skyrmion Hall angle, particularly in explaining its dependence on the skyrmion velocity, and this aspect is also analysed in this work in the following section. Another mechanism proposed is the effect of the field-like SOT [10], which is indeed quite significant as shown by Equation (10). The results in Figure 3 show that inclusion of diffusive spin torque can result in small skyrmion Hall angles even at moderate magnetisation damping values. We propose here the diffusive spin torque may help to explain the experimentally observed small skyrmion Hall angles and it is hoped these results will encourage further work in this direction.

**Figure 4** – Néel skyrmion motion in [Pt/Co/Ta]$_x$ disks (x = 1, …, 6) shown for $J_C = 1.3 \times 10^{11}$ A/m$^2$, D = -1.5 mJ/m$^2$ with $H_z$ = 15 kA/m. Results obtained using the self-consistent spin transport solver are shown with SHE (**ST**) and without SHE (**Diff.**). These results are contrasted with simulations using only the analytical form of the SOT. The effect of inter-layer spin diffusion on skyrmion motion is dependent on the number of Pt/Co/Ta repetitions, as is the SOT. The combined effect however saturates after 3 repetitions in this case (**ST**). The inset shows the calculated perpendicular effective spin polarisation parameter as a function of number of stack repetitions. The perpendicular effective non-adiabaticity parameter is not affected by the number of repetitions.

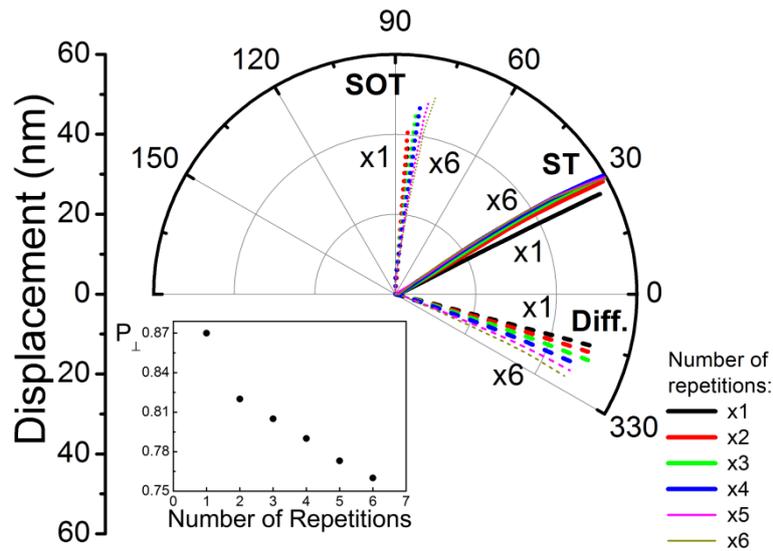

Increasing the number of stack repetitions results in modified demagnetising fields, and these are known to have a marked effect on skyrmion motion under a SOT [7,20]. Additionally, the total effective diffusive spin torque is affected by the number of repetitions in the multilayered stack. The spin accumulation generated at a skyrmion diffuses across the Pt and Ta layers, with the transverse components of the spin current absorbed in neighbouring Co layers. Due to the symmetry of the structure, a decrease in the overall diffusive spin torque is expected, reflected in a decrease of the effective perpendicular spin polarisation parameter $P_\perp$. The results for stacks with up to 6 number of repetitions are shown in Figure 4. With a single stack repetition the only contribution to the total effective diffusive spin torque is the main contribution due to diffusion from the Co layer into the adjacent Pt and Ta layers. From 2 stack repetitions up we have the additional contributions due to diffused spin currents between adjacent Co layers – with 2 stacks each Co layer has contributions due to the other Co layer only; with 3 stacks repetitions up, the inner Co layers experience contributions from the 2 adjacent layers. The total effect is a decrease in the total effective diffusive spin torque as the number of layers is increased – this is reflected in the decrease of $P_\perp$ as shown in the inset to Figure 4. It is interesting to note that as the number of repetitions is increased, the total spin torque, consisting of the combination of SOT and diffusive spin torque, results in the same skyrmion motion above 3 repetitions – thus for x = 4, 5, 6 the skyrmion paths and velocities are nearly identical, as seen in Figure 4. It must be stressed however this is not generally true since the skyrmion motion is a result of the interplay between the spin torques and demagnetising fields.

**Figure 5** – Single skyrmion velocities in [Pt/Co/Ta]$_x$ disks (x = 1, 2, 3) obtained as a function of current density for D = ±1.5 mJ/m$^2$ with H$_z$ = 15 kA/m, are shown as solid symbols together with linear fits. The skyrmion movement directions are shown as open symbols. Velocities obtained with spin pumping enabled are shown as triangles.

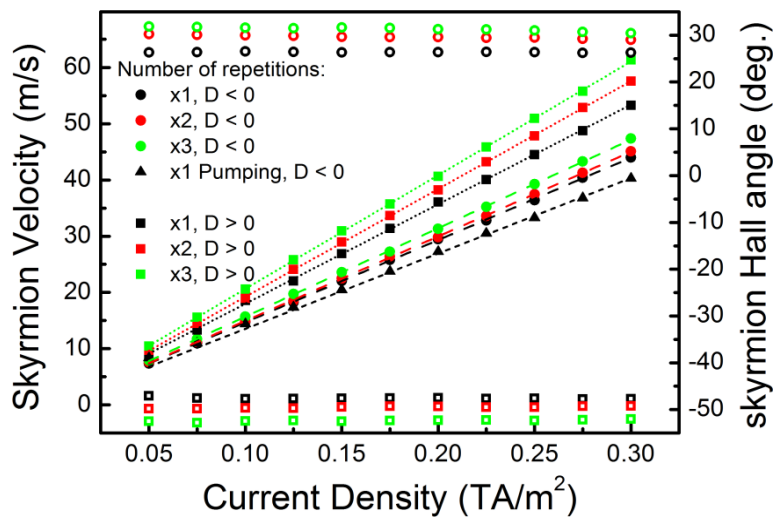

Finally, before analysing the effect of defects on skyrmion motion and threshold currents, the velocities are computed in perfect structures using the self-consistent spin transport solver. The results for up to 3 stack repetitions for both $D < 0$ and $D > 0$ are shown in Figure 5. Due to the symmetry of the various spin torques (see Figure 2) the velocities for $D > 0$ are greater than for $D < 0$, and moreover the motion in both cases opposes the drift direction of electrons. The skyrmion reaches its steady velocity almost instantaneously in these disk structures – any acceleration period is below the numerical error. The velocities obtained are very similar to those obtained in experiments on similar stack compositions with similar skyrmion diameters [7]. It must be stressed however that a precise comparison is difficult largely due to the unknown skyrmion Hall angle. Moreover the movement of skyrmions is also affected by the shape anisotropy of track structures, and disorder plays a very significant effect on the skyrmion movement path. Comparable skyrmion velocities have also been observed in other experimental studies [10-12], although the stack compositions are different. When spin pumping is also taken into account, a small decrease in velocity is observed in Figure 5 which is proportional to the driving current. This is explained as an increased spin drag effect as the skyrmion velocity increases, resulting in larger pumped spin current in Equation (9).

## IV. Threshold Currents

Experimental results on current-induced skyrmion motion show the existence of threshold currents required to initiate and sustain motion [7-12]. Further, the skyrmion Hall angle has been found to vary with skyrmion velocity [8-10]. These effects cannot be explained using simulations with perfect structures and constant material parameters. Material imperfections seem to play a very significant role in explaining these experimental observations. Previous studies have shown how a threshold current arises due to confining pinning potentials [32,33], polycrystalline structures with crystallites of varying anisotropy axis orientation [34], disorder originating from $M_S$ fluctuations using a granular structure [35], as well as disorder in the DMI [7,8] and anisotropy constant [7,11]. Imperfections have also been shown to result in a change of the skyrmion Hall angle with skyrmion velocity due to sliding motion along grain boundaries [34], effect of pinning defects due to anisotropy [11] and DMI [8] variation. Moreover Brownian motion of skyrmions due to thermal effects can result in distortions and diffusion of skyrmions [36,37].

Here we consider, in addition to variations of $M_S$ and $K_u$ parameters, also the effect of topographical surface roughness, included in simulations as a roughness field [38,39]. Topographical surface roughness results in an effective uniaxial anisotropy when averaged over the entire sample, however locally the roughness field has strong variations, which can result in confining potentials due to local fluctuations of the total effective anisotropy. Since the Co layers are very thin, variations in thickness of even a single monolayer can result in strong confining potentials. In this work we consider the effect of surface roughness up to 2 Å roughness per surface which is comparable to a single monolayer thickness variation. Roughness textures are generated as jagged granular profiles (see Methods section) with a 50 nm grain size, as shown in Figure 6 for topographical surface roughness. It is known that threshold currents depend on the skyrmion diameter to grain size ratio, with the strongest pinning obtained when this ratio is close to 1 [8]. Here we keep the grain size fixed in order to investigate threshold currents and variation of the skyrmion Hall angle. Further analysis using combinations of various sources of imperfections as well as grain size variation is outside the scope of the current work.

**Figure 6** – Effect of 2 Å surface roughness in a 320 nm diameter disk (plotted using color coding, ranging from blue at maximum depth of 2 Å to red), on skyrmion motion for various current densities, for both (a) 0 K, and (b) 297 K with thermal fluctuations. The skyrmion is initially at a confining site. Below the threshold current the skyrmion orbits within the confining site and is unable to escape. When the current is turned off the skyrmion returns to its initial position. Just above the threshold current the skyrmion escapes the confining site but its motion direction is strongly influenced by the local roughness profile, following a path with largest magnetic layer thickness. As the current density is increased the skyrmion path tends towards that obtained without surface roughness.

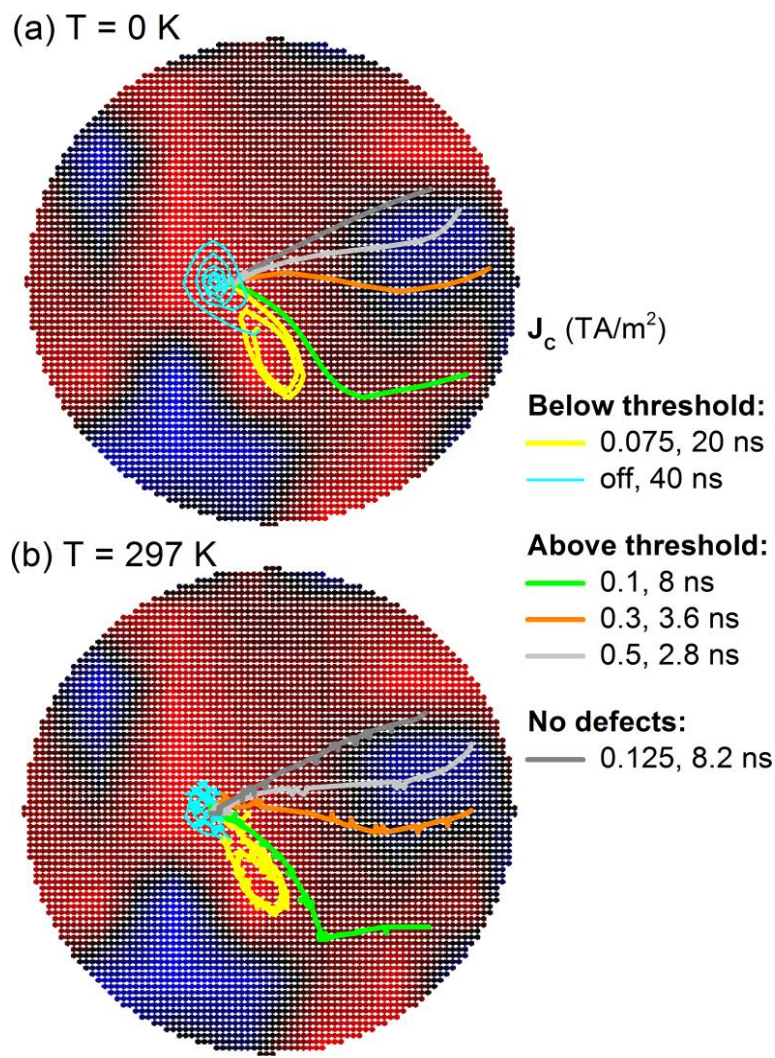

Results for skyrmion motion using surface roughness are shown in Figure 6, both for zero temperature and room temperature. For the latter a thermal field was also introduced as outlined in the Methods section. For the simulations in this section the skyrmion was relaxed into a confining site, then current densities of various strengths were applied and the skyrmion motion was computed using the full spin transport solver. For small current densities the skyrmion tends to undergo an orbiting motion inside the confining potential, as shown in Figure 6. As soon as the current is turned off, the skyrmion relaxes back to the initial position, representing the lowest energy configuration inside the two-dimensional confining potential. As the current is increased, eventually the skyrmion is able to escape. The calculated threshold current of $10^{11}$ A/m$^2$ for 2 Å surface roughness is very similar to that found in experiments [7]. The skyrmion motion is strongly influenced by the local roughness profile and can differ considerably from that obtained in perfect structures. With surface roughness the skyrmion tends to follow a path with greatest layer thickness, since this represents the lowest energy path. As the current density is increased the skyrmion path tends towards that obtained in perfect structures, as shown in Figure 6. Whilst the movement direction is strongly affected by the local roughness profile, the average skyrmion velocity above the threshold current is very similar to that obtained in perfect structures, as seen in Figure 7(a). When a stochastic thermal field is introduced for room-temperature simulations the skyrmion paths are largely unaffected, showing only a small random variation around the path taken without a thermal field. The threshold current is also unaffected. This suggests the additional Brownian motion of skyrmions is insufficient to overcome the pinning potentials in this case.

**Figure 7** – Skyrmion velocities and directions for various defect types, including $M_S$ and $K_u$ variations up to 15%, as well as surface roughness up to 2 Å per surface, showing (a) velocities as a function of current density, and (b), (c) average skyrmion movement direction as a function of current density.

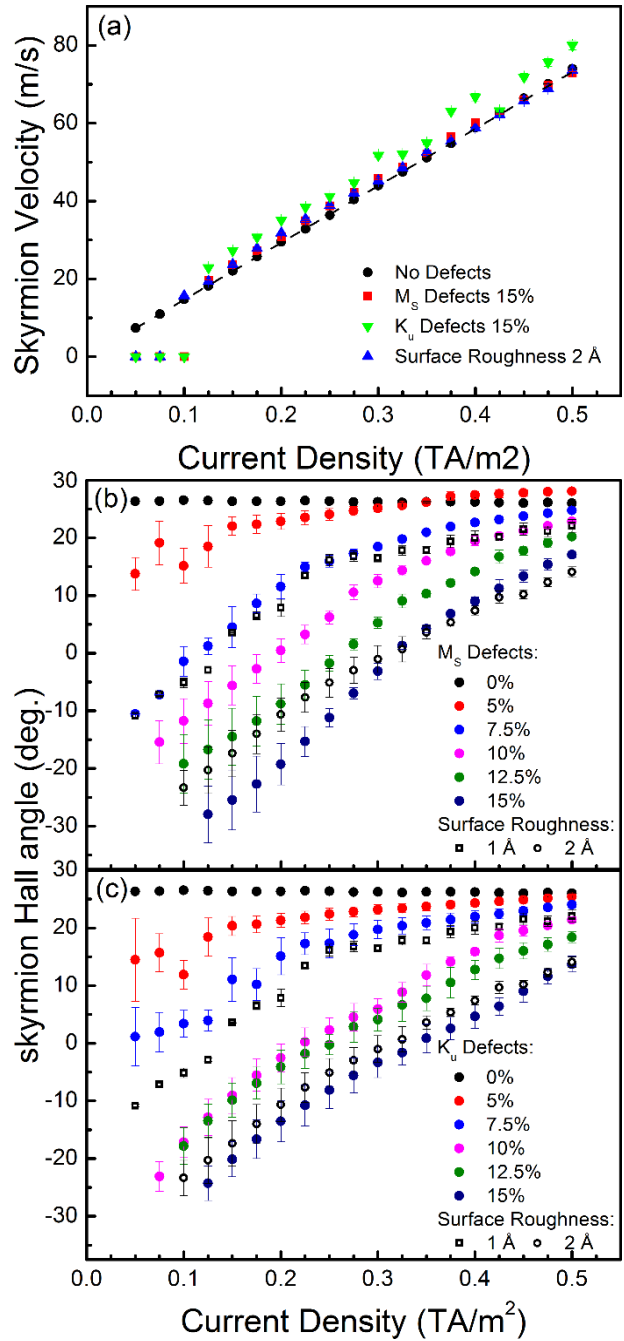

We further study the effect of magnetic defects, in particular considering variation of $M_S$ and $K_u$ parameters by changing the variation amplitude from 5% up to 15%. The results are shown in Figure 7. As expected, increasing the variation amplitude results in increasing threshold currents, with the largest threshold currents obtained for 15% variation as $1.25\times10^{11}$ A/m$^2$, comparable to that obtained for 2 Å surface roughness. It is unclear if such a strong parameter variation amplitude is likely in good quality samples, however a single monolayer variation at surfaces is, particularly in multilayered stacks considering the size of typical samples used to study skyrmion motion. The average skyrmion Hall angle is plotted in Figure 7 as a function of both current density and parameter variation amplitude. As the current density is increased the skyrmion Hall angle tends to that obtained in the ideal structure, levelling off as the current density is increased. Such a strong influence of the skyrmion velocity on its motion direction has also been observed in experimental studies [8-10]. Moreover, imaging of multiple skyrmions movement has shown simultaneously both negative and positive skyrmion Hall angles within the same driving current pulse [8]. The results in Figure 7 show how the sign of the skyrmion Hall angle can change depending on the level of disorder, as well as the driving current density, highlighting the effect local disorder can have on skyrmion movement.

Here we showed that in addition to magnetic defects, topographical surface roughness also plays a very important part. For spintronics devices using skyrmion motion it is important to have a skyrmion Hall angle of 0°, thus avoiding interactions with sharp sample boundaries which can result in annihilation of skyrmions. The results on surface roughness show it may be possible to design devices with skyrmion motion only along the track, by purposely enlarging the thickness of the track in the center. This creates a strong confining potential in the center of the track, whilst avoiding the sharp track boundaries, without significantly affecting the skyrmion speed. It is hoped these results will further stimulate experimental work in this direction.

## V. Conclusions

In conclusion, we have studied single skyrmion movement in ultra-thin multilayered Pt/Co/Ta disks by means of micromagnetics simulations coupled with a self-consistent spin transport solver. Vertical spin currents can drive skyrmions very efficiently in such structures. One source of vertical spin currents is the SHE, resulting in SOTs acting on the Co layers. Another source of vertical spin currents was shown here, resulting from inter-layer diffusion of a spin accumulation generated at a skyrmion. This diffusive spin torque was shown to act in the direction of electrical current irrespective of the skyrmion chirality or topological charge, and in ultra-thin films can be comparable to the SOT. The combination of SOT and diffusive spin torque was found to result in small skyrmion Hall angles even for small magnetisation damping values. Further, the effect of magnetic defects and topographical surface roughness on the skyrmion Hall angle and threshold current was studied. In particular topographical surface roughness, as small as a single monolayer variation, was shown to have a marked effect, resulting in a dependence of the skyrmion Hall angle on the skyrmion velocity, with threshold currents comparable to those found in experiments.

# VI. Methods

All simulations were done using Boris Computational Spintronics software, version 2.2 [40]. Material parameters used in the simulations are summarised in Table I.

**Table I** – Material parameters used to model the Pt/Co/Ta stacks.

| Parameter | Value | References |
|---|---|---|
| $|D|$ (Co) | 1.5 mJ/m$^2$ | [7] |
| $M_S$ (Co) | 600 kA/m | [7] |
| A (Co) | 10 pJ/m | [7] |
| $K_u$ (Co) | 380 kJ/m$^3$ | [7] |
| α (Co) | 0.03 up to 0.3 | [41] resp. [42] |
| $g_{rel}$ (Co) | 1.3 | [41] |
| σ (Co) | 5 MS/m | [43] |
| σ (Pt) | 7 MS/m | [44] |
| σ (Ta) | 500 kS/m | [45] |
| P (Co) | 0.4 | [46,47] |
| De (Co, Pt, Ta) | 0.01 m$^2$/s | [43] |
| $λ_{sf}$ (Co) | 38 nm | [43, 48] |
| $λ_{sf}$ (Pt) | 1.4 nm | [44] |
| $λ_{sf}$ (Ta) | 1.9 nm | [45] |
| $λ_J$ (Co) | 2 nm | [18] |
| $λ_φ$ (Co) | 4 nm | [49, 50] |
| $G^{↑↓}$ (Pt/Co) | 1.5 PS/m$^2$ | [44] |
| $G^{↑↓}$ (Co/Ta) | 1 PS/m$^2$ | [45] |
| $θ_{SHA}$ (Pt) | 0.19 | [44,27] |
| $θ_{SHA}$ (Ta) | -0.15 | [27,51] |

In Table I the spin dephasing length is given is given by $\lambda_\varphi = \lambda_J \sqrt{l_\perp / l_L}$, where $l_\perp$ and $l_L$ are the transverse spin coherence and spin precession lengths respectively [49, 50], estimated as 4 nm for Co.

Computations were done using cell-centered finite difference discretisation. Differential operators are evaluated to second order accuracy in space, for both magnetisation and spin transport calculations. For magnetisation dynamics the computational cellsize used was (4 nm, 4nm, 1 nm). For spin transport calculations the computational cellsize used was (4 nm, 4 nm, 0.5 nm) for the Pt and Ta layers, and (4 nm, 4 nm, 0.25 nm) for the Co layers. The LLG equation was evaluated using the RK4 evaluation with a 0.5 ps fixed time step. The Poisson equations for spin and charge transport, e.g. Equation (4) for **S**, were evaluated using the successive over-relaxation algorithm. All computations were done on the GPU using the CUDA C framework.

In the LLG equation the contributing interactions are the demagnetising field, direct exchange interaction, the interfacial Dzyaloshinskii-Moriya exchange interaction, uniaxial magneto-crystalline anisotropy and applied field. The roughness field resulting from topological surface roughness is described in [38]. Roughness profiles were generated using a jagged granular generator algorithm. Equally spaced coefficients at 50 nm spacing in the x-y plane are randomly generated. The remaining coefficients are obtained using bi-linear interpolation from the randomly generated points. The resulting array of coefficients in the x-y plane are used to locally multiply the base parameter values, $M_S$ and $K_u$, or to obtain a topographical surface roughness profile – see Figure 6. Further details are given in the user manual for Boris [40]. Simulations with a thermal field at room temperature were done using the stochastic LLG equation, evaluated using Heun's method, with a thermal field obtained as:

$$|\mathbf{H}_{thermal}|_{max} = \sqrt{\frac{2k_B T}{\alpha \gamma \mu_0 M_s^0 V \Delta t}},$$

where $V$ is the computational cell volume, $\Delta t$ is the time step, and $T$ is the temperature.

# Effect of inter-layer spin diffusion on skyrmion motion in magnetic multilayers

## Supplementary Material

Serban Lepadatu

*Jeremiah Horrocks Institute for Mathematics, Physics and Astronomy, University of Central Lancashire, Preston PR1 2HE, U.K.*

The inter-layer diffusive spin torque has the form:

$$\mathbf{T}_{Diff.} = -\left[ (\mathbf{u}_\perp.\nabla)\mathbf{M} - \frac{\beta_\perp}{M_S}\mathbf{M}\times(\mathbf{u}_\perp.\nabla)\mathbf{M} \right] \qquad (S1)$$

This was verified in the main text by comparison of full spin transport solver simulations with simpler simulations based on the LLG equation augmented by the torque in Equation (S1). To see directly how this torque arises we consider the drift-diffusion model for a simple N/F bilayer without the spin-Hall effect. For definitions of parameters see the main text. Within this model, the spin current crossing the N/F interface is absorbed as given by:

$$\mathbf{J}_S.\mathbf{n}\big|_N - \mathbf{J}_S.\mathbf{n}\big|_F = 2D_e\left[ \text{Re}\{G^{\uparrow\downarrow}\}\mathbf{m}\times\mathbf{m}\times\left(\frac{\mathbf{S}_F}{\sigma_F} - \frac{\mathbf{S}_N}{\sigma_F}\right) + \text{Im}\{G^{\uparrow\downarrow}\}\mathbf{m}\times\left(\frac{\mathbf{S}_F}{\sigma_F} - \frac{\mathbf{S}_N}{\sigma_F}\right) \right] \qquad (S2)$$

The longitudinal component of the spin current flowing perpendicular to the interface (z direction) inside the F layer is proportional to the z-direction derivative of the spin accumulation. This is negligible due to the long spin diffusion length in Co, as shown in Figure S1 where the spin accumulation is plotted perpendicular to the interface for a Pt/Co bilayer containing a skyrmion. On the other hand the transverse spin current inside the N layer is not negligible, and arises due to the decay (diffusion) of the spin accumulation generated in the F layer. This results in an interfacial spin torque given by:

$$\mathbf{T}_S = \frac{gD_e}{d_F}\left[\operatorname{Re}\{G^{\uparrow\downarrow}\}\mathbf{m}\times\mathbf{m}\times\left(\frac{\mathbf{S}_F}{\sigma_F}-\frac{\mathbf{S}_N}{\sigma_F}\right)+\operatorname{Im}\{G^{\uparrow\downarrow}\}\mathbf{m}\times\left(\frac{\mathbf{S}_F}{\sigma_F}-\frac{\mathbf{S}_N}{\sigma_F}\right)\right] \quad (S3)$$

**Figure S1** – Spin accumulation for a Pt/Co bilayer without the spin-Hall effect, plotted along the z direction through the center of a skyrmion.

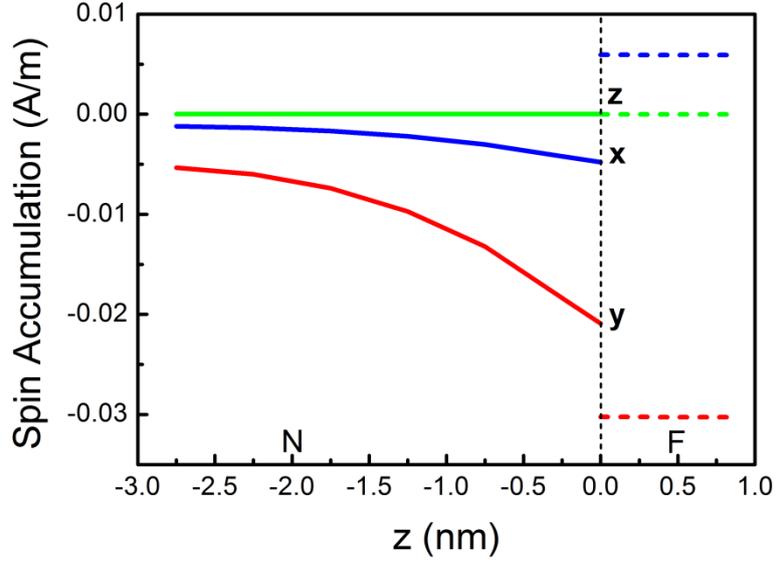

To calculate this, we need to obtain expressions for $\mathbf{S}_F$ and $\mathbf{S}_N$, the spin accumulations on either side of the interface. The spin accumulation in the F layer satisfies the following equation of motion:

$$\frac{\partial \mathbf{S}}{\partial t} = P\frac{\mu_B}{e}(\mathbf{J}_C\cdot\nabla)\mathbf{m} - D_e\left(\frac{\mathbf{S}}{\lambda_{sf}^2}+\frac{\mathbf{S}\times\mathbf{m}}{\lambda_J^2}+\frac{\mathbf{m}\times(\mathbf{S}\times\mathbf{m})}{\lambda_\varphi^2}\right)+D_e\nabla^2\mathbf{S} \quad (S4)$$

We can solve Equation (S4) for the steady state to obtain the spin accumulation in the F layer. To do this we first ignore in-plane spin diffusion, i.e. $\nabla^2\mathbf{S} \cong 0$, thus obtaining the spin accumulation on the F side as:

$$\mathbf{S}_F = a_F(\mathbf{u}.\nabla)\mathbf{M} + \frac{b_F}{M_S}\mathbf{M}\times(\mathbf{u}.\nabla)\mathbf{M} \tag{S5}$$

Here $a_F$ and $b_F$ are given by:

$$a_F = \frac{\lambda_{sf,F}^2}{D_e}\frac{a^2+bc}{a^2+c^2}$$

$$b_F = \frac{\lambda_{sf,F}^2}{D_e}\frac{a^3+ab^2}{a^2+c^2} \tag{S6}$$

with

$$a = \frac{\lambda_{sf,F}^2}{\lambda_J^2}, \quad b = \frac{\lambda_{sf,F}^2}{\lambda_\varphi^2}, \quad c = a^2+b+b^2 \tag{S7}$$

The spin accumulation in Equation (S5) leads to the Zhang-Li spin-transfer torque [1,2] in the plane. In multi-layers this same spin accumulation results in vertical spin currents due to inter-layer diffusion and an interfacial spin torque as illustrated in Figure S1, which scales inversely with the F layer thickness. On the N side, without the spin-Hall effect the vertical spin current is simply proportional to the z-direction derivative of the spin accumulation, where again we ignore any in-plane spin accumulation gradients. With the composite media boundary condition in Equation (S2) and the requirement of zero spin current normal to the sample boundary, we obtain the spin accumulation on the N side of the interface as:

$$\mathbf{S}_N = a_N(\mathbf{u}.\nabla)\mathbf{M} + \frac{b_N}{M_S}\mathbf{M}\times(\mathbf{u}.\nabla)\mathbf{M} \tag{S8}$$

Here we have in turn:

$$a_N = \frac{(1+d)d_+ + dd_-}{2d^2+2d+1}$$

$$b_N = \frac{dd_+ - (1+d)d_-}{2d^2+2d+1} \tag{S9}$$

with

$$d = \frac{\tilde{G}_N^{\uparrow\downarrow}\lambda_{sf,N}}{\tanh\left(\frac{d_N}{\lambda_{sf,N}}\right)}, \quad d_\pm = \frac{\tilde{G}_F^{\uparrow\downarrow}\lambda_{sf,N}}{\tanh\left(\frac{d_N}{\lambda_{sf,N}}\right)}(a_F \pm b_F) \tag{S10}$$

where $\tilde{G}_x^{\uparrow\downarrow} = 2\text{Re}\{G^{\uparrow\downarrow}\}/\sigma_x$ $(x=N,F)$, and $d_N$ is the N layer thickness.

Using Equations (S5) and (S8) in Equation (S3), Equation (S1) results, where we have:

$$\mathbf{u}_\perp = \mathbf{J}_C \frac{P_\perp g \mu_B}{2 e M_S}, \tag{S11}$$

The effective parameters $P_\perp$ and $\beta_\perp$ are given as:

$$P_\perp = P(A+B)$$
$$\beta_\perp = (A-B)/(A+B) \tag{S12}$$

where

$$A = \frac{2 D_e \operatorname{Re}\{G^{\uparrow\downarrow}\}}{d_F} \left( \frac{a_F}{\sigma_F} - \frac{a_N}{\sigma_N} \right)$$

$$B = \frac{2 D_e \operatorname{Re}\{G^{\uparrow\downarrow}\}}{d_F} \left( \frac{b_F}{\sigma_F} - \frac{b_N}{\sigma_N} \right) \tag{S13}$$

The above derivation shows how the diffusive spin torque in Equation (S1) arises for an N/F bilayer under the simplifying assumption of negligible in-plane diffusion. For more complicated multilayered structures, corrections to the $P_\perp$ and $\beta_\perp$ parameters are required, however the form of the diffusive spin torque remains the same. With in-plane spin diffusion taken into account a modified in-plane spin accumulation results in the F layer [3]. The self-consistent spin transport solver used in this work takes into account all these effects, including in-plane spin diffusion in the N layers, combining both the SOT and diffusive spin torque into a single interfacial spin torque included in simulations. The disadvantage of this approach is increased simulation time due to the additional iterations required to solve the Poisson-type equations for the drift-diffusion model. This can increase the computation time by 5 times or more, thus simpler simulations using just the LLG equation augmented with the SOT and the diffusive spin torque in Equation (S1) are attractive.

Note, the current version of the micromagnetics software used for this work [4] employs the successive over-relaxation algorithm with black-red ordering for parallelization to solve Poisson equations. Whilst this method is robust and able to handle arbitrary multi-layered structures and shapes, it does suffer from slow convergence for lower target solver errors. It is expected a more efficient algorithm, including the alternating direction iterative method with parallelized Thomas algorithm, a FFT-based Poisson solver, or a bi-conjugate gradient method, will significantly improve computation time in a future version.

**Figure S2** – Diffusive interfacial spin torque for a Pt/Co/Ta stack with skyrmions of different diameters as indicated, obtained using the spin transport solver and compared with Equation (S1) for $P_\perp = 0.87$ and $\beta_\perp = -0.132$.

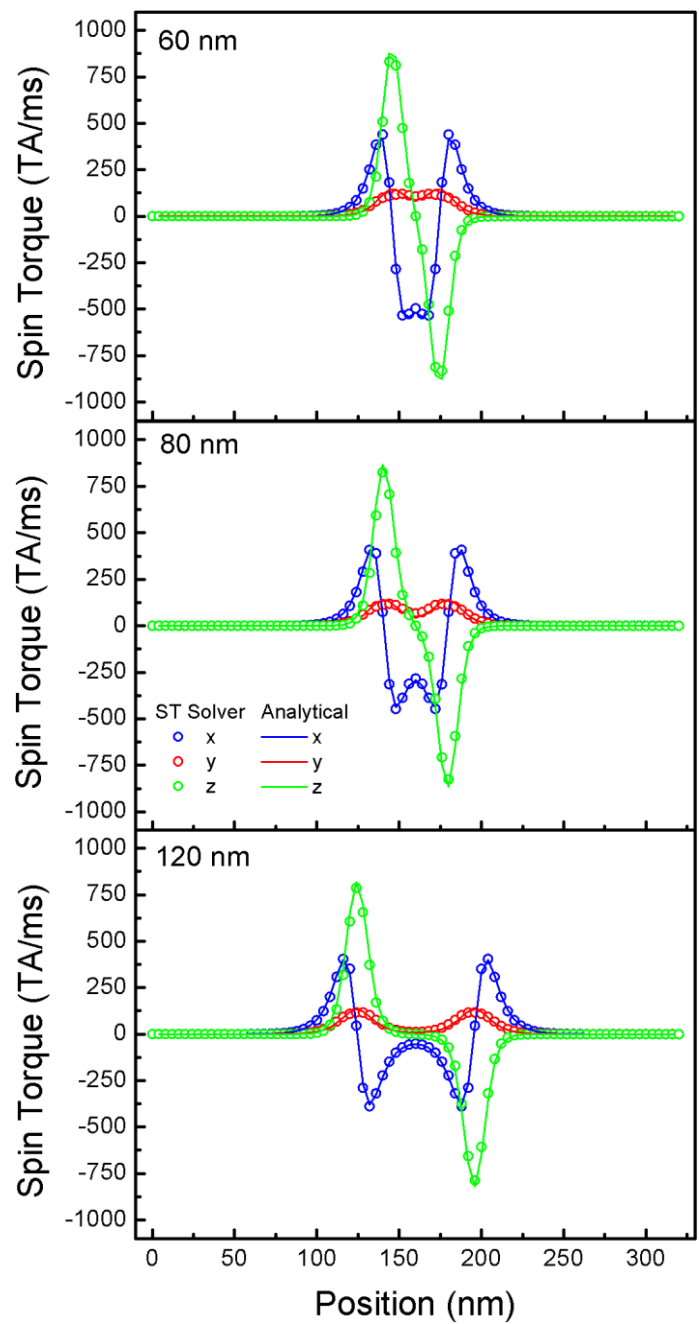

To use Equation (S1) in simulations without the spin-transport solver, the $P_\perp$ and $\beta_\perp$ parameters should be obtained for the multi-layered structure under study. The $P_\perp$ and $\beta_\perp$ parameters are effective parameters for the entire stack and may be obtained readily by fitting to computed diffusive spin torques with Equation (S1). This is shown in Figure S2 for 3 different skyrmion diameters.

## Supplementary Material References